\documentclass[12pt]{article}
\usepackage{epsfig}
\usepackage{amsmath}
\def\beq{\begin{equation}}
\def\eeq#1{\label{#1}\end{equation}}
\def\eeqn{\end{equation}}

\def\beqa{\begin{eqnarray}}
\def\eeqa#1{\label{#1}\end{eqnarray}}
\def\eeqan{\end{eqnarray}}

\let\bar=\overbar

\def\Dslash{\not{\hbox{\kern-4pt $D$}}}
\def\dslash{\not{\hbox{\kern-2pt $\del$}}}

\def\msb{{\bar{\ssstyle M \kern -1pt S}}}

\usepackage{fancyhdr,graphicx}
\def\tstrut{\vrule height2.5ex depth0pt width0pt} 
\def\Title#1{\begin{center} {\Large {\bf #1} } \end{center}}

\begin{document}

\Title{$\Lambda\Lambda$ interaction and hypernuclei}

\bigskip\bigskip


\begin{raggedright}

{\it C. Albertus$^1$, J. E. Amaro$^1$, J. Nieves$^2$\\ $^1$ Departamento
  de F\'\i sica At\'omica, Molecular y Nuclear. Facultad de
  Ciencias. Universidad de Granada.\\ Avenida de Fuentenueva S/N,
  E-18071, Spain\\ $^2$ Instituto de F\'\i sica Corpuscular (IFIC),
  Centro Mixto CSIC-Universidad de Valencia, Institutos de
  Investigaci\'on de Paterna.\\ Apartado 22085, E-46071 Valencia,
  Spain\\ {\tt Email: albertus@ugr.es}}
\bigskip\bigskip
\end{raggedright}

\section{Introduction}

Whether the existence of hyperonic matter is present in the central regions of neutron stars with densities are in excess of nuclear saturation density is unknown. One of the fundamental pieces to determine this is the interaction among baryons bearing strangeness.  A neutron star has been considered as a gigantic nucleus of $N\sim 10^{58}$ nucleons ~\cite{glen} with a radial structure closely related to the equation of state of nuclear matter and, in turn, to the nuclear interaction. Due to the large complexity of the interior of this macroscopic object with typical radii of $\sim12$ km and masses of $1.5M_\odot$ is not possible to obtain an in-detail description so far. However we can consider a sort of replica and helpful insight from finite nuclei orders of magnitude much smaller. Even, from the existence of hypernuclei, nuclei where part of its content consists of hyperons. In this sense non-relativistic \cite{nr} as well as relativistic field models have been developed (and continue to present) in the literature \cite{rel} to describe the phenomenology.

In particular, because of the lack of data from scattering experiments,
$\Lambda\Lambda$ hypernuclei provide a valuable method to learn
details on the baryon-baryon interaction in the strangeness $S=-2$
sector. Furthermore, in the last years a considerable effort has been
done, both by the experimental and theoretical communities, in the
physics of single and double $\Lambda$ hypernuclei. Many single
$\Lambda$ hypernuclei and some double $\Lambda$ hypernuclei have been
observed and their energies have been measured.

In this work we present the model of Ref.~\cite{aan02} to calculate the
binding energy of double $\Lambda$ hypernuclei, defined as
\begin{equation}
B_{\Lambda\Lambda}=-[M(^{A+2}_{\Lambda\Lambda}Z)-M(^AZ)-2m_{\Lambda}].
\end{equation}
In what follows, we will outline the main features of the scheme of
reference \cite{aan02} to account for the modification of the
interaction between the two $\Lambda$ baryons by the presence of the
nuclear core.

\section{Model for $\Lambda\Lambda$ hypernuclei}

We model a $\Lambda\Lambda$ hypernucleus as an interacting pair of
$\Lambda$ hyperons plus a nuclear core. We adopt a variational
approach to determine the intrinsic wave function and to calculate the
binding energies.

Once we have removed the center of mass, we write the intrinsic Hamiltonian as
\begin{align}
H&=h_{sp}(1)+h_{sp}(2)+V_{\Lambda\Lambda}(1,2)-\frac{\nabla_1\cdot\nabla_2}{M_A}
\nonumber\\
h_{sp}&=-\frac{\nabla^2_i}{2\mu_A} + V_\Lambda(|\vec r_i|)
\label{eq:hamiltonian}
\end{align}
where $\mu_A$ and $M_A$ are the reduced mass of the $\Lambda$
hyperon-nucleus system and the mass of the nuclear core, respectively.

The $V_\Lambda$ potential in the single particle Hamiltonian $h_{sp}$
accounts for the $\Lambda$ nucleus interaction and has been adjusted
to reproduce the binding energy $B_\Lambda =
-[M(_\Lambda^{A+1}Z)-M(^AZ)-m_\Lambda]$ of the corresponding single
$\Lambda$ hypernuclei. $V_{\Lambda\Lambda}$ represents the interaction
between the two $\Lambda$ hyperons in the nuclear medium. The presence
of a second $\Lambda$ hyperon results in a dynamical reordering of the
nuclear core. This reordering effect in the nuclear core and the free
space $\Lambda\Lambda$ interaction itself contributes to $\Delta
B_{\Lambda\Lambda}=B_{\Lambda\Lambda}-2B_\Lambda$, although the former
effect is suppressed compared to the latter by, at least, by one power
of the nuclear density. We have assumed that this nuclear core
dynamical reordering effect amounts to be around $0.5$ MeV for light
$\Lambda\Lambda$-hypernuclei, as suggested by the $\alpha-$cluster
model calculations, and negligible for heavy ones. This uncertainty is
of the order of the experimental errors~\cite{buc84}. One should
mention as well that this reordering effect is also partially taken
into account in the RPA calculation described below.

\section{Free space $\Lambda\Lambda$ interaction}

We use Bonn-J\"ulich models to construct the free space
$\Lambda\Lambda$ interaction. We consider the exchange of $\sigma$
($I=0, J^p=0^+$), $\omega$ and $\phi$ ($I=0, J^P= 1^-$) mesons between
the two $\Lambda$ hyperons. Furthermore, monopolar form factors are
used, leading to extended expressions for the potentials. Once the
form factors are included, the potentials turn out to be:
\begin{align}
V_{\sigma}(r)&=-m_{\sigma}\frac{g_{\sigma\Lambda\Lambda}^{2}}{4\pi}
\left\{\tilde{Y}(\sigma,r)+
\frac{1}{2m^{2}_{\Lambda}}
\left[\left(\vec\nabla\tilde{Y}(\sigma,r)\right)
\cdot\vec\nabla+
\tilde{Y}(\sigma,r)\vec\nabla\,^{2}\right]\right\}\nonumber\\
V_{\alpha}(r)&=\frac{m_{\alpha}}{4\pi}\left\{
\hat{g}_{\alpha\Lambda\Lambda}^{2}\tilde{Y}(\alpha,r)+
\frac{g^{2}_{\alpha\Lambda\Lambda}-\hat{g}_{\alpha\Lambda\Lambda}^{2}}{m_{\alpha}^{2}}
\frac{(\Lambda_{\alpha\Lambda\Lambda}^{2}-m^{2}_{\alpha})^{2}}{2m_{\alpha}\Lambda_{\alpha\Lambda\Lambda}}
e^{-\Lambda_{\alpha\Lambda\Lambda}r}\right.- \nonumber \\
&\left.-\frac{3g^{2}_{\alpha\Lambda\Lambda}}{2m_{\Lambda}^{2}}
\left[\left(\vec\nabla\tilde{Y}(\alpha,r)\right)
\cdot\vec\nabla+
\tilde{Y}(\alpha,r)\vec\nabla\,^{2}\right]
\right\},\hspace{2ex}\alpha=\omega,\phi
\label{eq:pots1}
\end{align}
with
\begin{align}
\hat{g}_{\alpha\Lambda\Lambda}^{2}&=g^{2}_{\alpha\Lambda\Lambda}-
\frac{1}{2}\left(\left(\frac{m_{\alpha}}{m_{\Lambda}}\right)^{2}
\frac{3g_{\alpha\Lambda\Lambda}^{2}}{2}+\frac{m_{\Lambda}}{m_{N}}g_{\alpha\Lambda\Lambda}f_{\alpha\Lambda\Lambda}+
\left(\frac{m_{\Lambda}f_{\alpha\Lambda\Lambda}}{m_{N}}\right)^{2}\right)\nonumber\\
%
\tilde{Y}(\alpha,r)&=Y(m_{\alpha}r)-\left\{1+
\frac{r}{2\Lambda_{\alpha\Lambda\Lambda}}\left(\Lambda_{\alpha\Lambda\Lambda}^{2}-m_{\alpha}^{2}\right)\right\}
\frac{\Lambda_{\alpha\Lambda\Lambda}}{m_{\alpha}}Y\left(\Lambda_{\alpha\Lambda\Lambda}r\right)\nonumber\\
Y(x)&=\frac{e^{-x}}{x}.
\end{align}
In the above expressions $\alpha$ stands for $\omega$ and $\phi$.  The
value of the couplings constants and the cutoff
masses~\cite{caro99,reuber94} are summarized in
Table~\ref{tab:couplings}.

Bonn-J\"ulich models use the $SU(6)$ symmetry to relate the coupling
constants of the $\omega$ and $\phi$ mesons to the $\Lambda$ hyperons
to those of these mesons to the nucleons. Besides, we adopt the so
called "ideal mixing" and consider that the $\phi$ meson is a $s\bar
s$ state, leading to a vanishing $g_{\phi NN}$ coupling
constant. Hence, the $g_{\phi\Lambda \Lambda}$ coupling constant is
determined from $g_{\omega \Lambda\Lambda}$. Besides, as the $\phi$
meson does not couple to nucleons, exists a larger uncertainty in the
value of its cutoff. We assume for $\Lambda_{\phi\Lambda\Lambda}$ a
value similar or greater than $\Lambda_{\omega\Lambda\Lambda}$,
considering finally three values for $\Lambda_{\phi\Lambda\Lambda}$:
$1.5$, $2$ and $2.5$ GeV.

\begin{table}[b]
\begin{center}
\begin{tabular}{cccc}
\hline
\tstrut
Vertex & $g_\alpha/\sqrt{4\pi}$ & $f_\alpha/\sqrt{4\pi}$ & $\Lambda_\alpha$ (GeV)\\
\hline
$\omega\Lambda\Lambda$ & 2.981   & -2.796                 & 2 \\
$\sigma\Lambda\Lambda$ & 2.138   & -                      & 1 \\
$\phi\Lambda\Lambda$   & -2.108  & -3.954                 & 1.5--2.5 \\
\hline
\end{tabular}
\caption{\small \label{tab:couplings} \small Coupling constants and
  cutoffs used in Eq.~(\ref{eq:pots1}). These values have been taken
  from model $\hat{A}$ of Ref.~\cite{reuber94} }
\end{center}
\end{table}

\section{In medium contribution}
\begin{figure}[htb]
\begin{center}
\epsfig{file=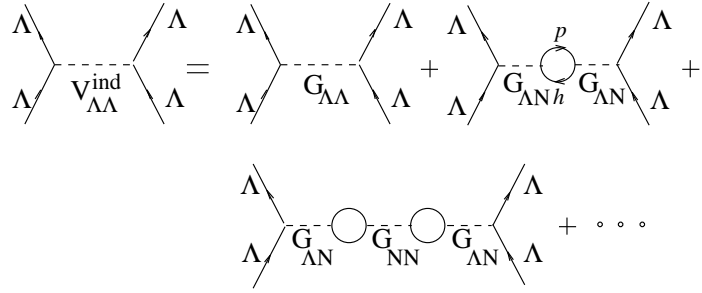,height=1.5in}
\caption{Diagrammatic definition of $V^{\rm ind}_{\Lambda\Lambda}$}
\label{fig:rpasum}
\end{center}
\end{figure}

Now we describe the summation of the diagrams of
Fig.~\ref{fig:rpasum}. We will do it first in nuclear matter and then
will apply that results to finite nuclei.

To evaluate the series of diagrams of Fig.~\ref{fig:rpasum} in
nuclear matter, we consider the case of a noninteracting Fermi gas of
nucleons with density $\rho$. The series of diagrams we are interested
in is just a diagrammatic representation of a Dyson type equation,
which modifies the propagation in nuclear matter of the carriers
($\sigma$, $\omega$ and $\phi$) of the interaction. As explained
before, the $\phi$ meson does not couple to nucleons and thus, its
propagation is not modified in the nuclear medium. The $\sigma-\omega$
propagator in the medium, is determined by the Dyson equation
\begin{equation}
D(Q) = D^0(Q)+D^0(Q)\Pi(Q)D(Q)
\label{eq:Dyson}
\end{equation}
\noindent where $D^0$ is the $5 \times 5$ matrix composed of the free
$\sigma$ and $\omega$ propagators,
\begin{equation}
D^0(Q)=\left[\begin{array}{cc}
D^{\omega}_{\mu\nu}(Q)  &   0\\
0                              &   D^{\sigma}(Q)
\end{array}\right],
\end{equation}

\noindent and $\Pi$ is the $\sigma-\omega$ self-energy in the nuclear
medium,

\begin{equation}
\Pi(Q)=\left[\begin{array}{cc}
\Pi(Q)_{\mu\nu}         &   \Pi(Q)_{\mu}\\
\Pi(Q)_{\nu}            &   \Pi(Q)_{s}
\end{array}\right],\label{eq:auto}
\end{equation}

\noindent where $\Pi_{\mu\nu}$ and $\Pi_s$ accounts for excitations
over the Fermi sea driven by the $\omega$ and $\sigma$ mesons,
respectively and $\Pi_\mu$ generates mixing for scalar and vector meson
propagation in the medium. This term vanish in the vacuum.

\begin{figure}[htb]
\begin{center}
\epsfig{file=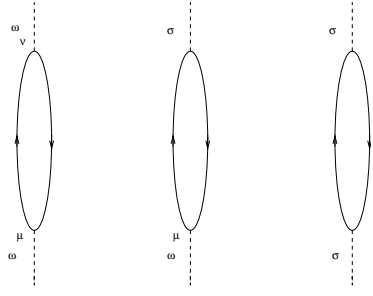,height=1.5in}
\caption{$p-h$ excitations contributing to $\Pi$.}
\label{fig:burbujas}
\end{center}
\end{figure}

We assume the following approximations in the evaluation of $\Pi(Q)$:
(i)We approximate $G_{\Lambda N}$ and $G_{NN}$ by the free space
diagonal $\Lambda N$ and $NN$ potentials, well described by $\sigma$
and $\omega$ exchanges in the isoscalar $^1S_0$ channels. The
$\Lambda\Lambda\sigma$ and $\Lambda\Lambda\omega$ vertices were
discussed above and the $NN\sigma$ and $NN\omega$ Lagrangians and
coupling constants can be found in \cite{machleidt87}. (ii) We have only
considered $p-h$ excitations over the Fermi sea. This amounts to
evaluate the diagrams of Fig.~\ref{fig:burbujas}, plus the corresponding
crossed terms. (iii) We have worked in a nonrelativistic Fermi sea and
evaluate the $p-h$ excitations in the static limit.

With all these approximations and taking the transferred four-momentum
between the two $\Lambda$ hyperons as $Q=(q^0=0,0,0,q)$, the elements of the
matrix $\Pi(0,q)$ can be written as:
\begin{align}
\Pi_{ij}(0,q)&=U(0,q;\rho)C_i^N(q)C_j^N(q);\hspace{1cm} i,j=1,\ldots,5\nonumber\\
C^B&\equiv (g_{\omega BB}(q),0,0,0,g_{\sigma BB}(q))\nonumber\\
g_{\alpha BB}(q) &= g_{\alpha BB}\frac{\Lambda_{\alpha BB}^2-m_{\alpha}^2}{\Lambda_{\alpha BB}^2+q^2} 
\label{eq:autopi}
\end{align}
where $\alpha = \omega,\sigma$ and $B=\Lambda , N$. In the Lindhard
function $U(0,q;\rho)$ a finite excitation gap has been included for
particles. We have used values for the gap between 1 and 3 MeV, to
account for typical excitation energies in finite nuclei. In the case
of $^4{\rm He}$, we used a value of 20 MeV for the gap. With
Eq.~(\ref{eq:autopi}), we can invert the Dyson equation, and one gets
\begin{equation}
D(Q)=(I-D^0(Q)\Pi(Q))^{-1}D^0(Q).
\end{equation}
With this propagator, the RPA series of diagrams of Fig.~\ref{fig:rpasum} (from
the second diagram on) can be evaluated and the RPA contribution to
the $\Lambda\Lambda$ interaction in nuclear matter results to be
\begin{eqnarray}
\delta V_{\Lambda\Lambda}^{RPA} (q,\rho) &=& \sum_{ij=1}^5 C_i^\Lambda(q) \left
[ D(Q)-D^0(Q)\right]_{ij} C_j^\Lambda(q) \nonumber\\
&=&
U(0,q;\rho) \frac{\left(W_{\Lambda N}^\sigma - W_{\Lambda
N}^\omega\right)^2}{1+ U\left(W_{NN}^\sigma -
W_{NN}^\omega\right)}\label{eq:solrpa}
\end{eqnarray}
where we have subtracted $D^0(Q)$ to avoid double counting and we have
defined $W^{\alpha}_{BB'}=\frac{g_{\alpha BB }(q)g_{\alpha
    B'B'}(q)}{q^2+m_\alpha^2}$.  We have neglected in this
nonrelativistic approach the spatial and tensor ($f_{\omega
  \Lambda\Lambda}$) couplings of the $\omega$ meson to the $\Lambda$.
$\delta V_{\Lambda\Lambda}^{RPA}(r_{12},\rho)$ depends on the distance
between the two $\Lambda$'s, $r_{12}$ and the constant density
$\rho$. In the case of a finite nuclei, the carrier of the interaction
feels different densities when it is travelling from one hyperon to
the other. To account for this fact in the case of finite nuclei, we
average over the densities the carrier feels along its flight. We
assume meson straight line trajectories and the local density
approximation, thus we obtain
\begin{equation}
\delta V_{\Lambda\Lambda}^{RPA}(1,2) = \int_{0}^{1}d\lambda 
\delta V_{\Lambda\Lambda}^{RPA}(r_{12},\rho(\left|\vec{r}_2+\lambda\vec{r}_{12}\right|))
\end{equation}
where $\rho$ is the nucleon center density given in Table~4 of
Ref.~\cite{caro99}.

\section{Variational approach}

We take advantage of the Variational Theorem to find the energy of the
ground state of the Hamiltonian of Eq.~(\ref{eq:hamiltonian}).

We have used a family of $^1S_0$ $\Lambda\Lambda$ wave functions of
the form $\Phi_{\Lambda\Lambda}(\vec r_1,\vec r_2) =
NF(r_{12})\phi_{\Lambda}(r_1)\phi_\Lambda(r_2)\chi^{S=0}$, with
$\chi^{S=0}$ the spin singlet. $N$ is a normalization constant and
$\vec r_{12}=\vec r_1 -\vec r_2$. The functions $\phi_\Lambda(r_i)$
are exact solutions of the single particle Hamiltonian
$h_{sp}$. $F(r_{12})$ is a Jastrow correlation function of the form
\begin{equation}
F(r_{12}) =\left(1+\frac{a_{1}}{1+(\frac{r_{12}-R}{b_{1}})^{2}}\right)
\prod_{i=2}^3\left(1-a_{i}e^{-b_{i}^2 r_{12}^2}\right)\label{eq:var2}
\end{equation}
where $a_{i},b_i,R,\hspace{2mm}i=1,3\hspace{2mm}$ are free
parameters. The values of the parameters for which the expected value
of the Hamiltonian reaches a minimum are summarized in Table II
of Ref.~\cite{aan02}.

\section{Results and discussion}

Using the $\Lambda$ nuclear core potentials summarized in
Ref.~\cite{caro99}, together with the $\Lambda\Lambda$ interaction and
the variational wave functions described above, we obtain the
results~\cite{aan02} of Table~\ref{tab:results}. We have also
considered the dependence of the results on the couplings, by varying
them $\pm 10\%$ around their SU(6) values, finding appreciable
variations.

In Table~\ref{tab:results}, the experimental values for the binding
energy of $^{\phantom{6}6}_{\Lambda\Lambda}$He and
$^{13}_{\Lambda\Lambda}$B\cite{nakazawa10}, are more updated with
respect to those included in Ref.~\cite{aan02}. Ref.~\cite{nakazawa10}
reports a value of $B_{\Lambda\Lambda}$($^{10}_{\Lambda\Lambda}$Be) =
11.90 MeV, much smaller than those reported in the same reference for
$^{11}_{\Lambda\Lambda}$Be and $^{12}_{\Lambda\Lambda}$Be, 20.49 and
22.23 MeV respectively, for which the nuclear cores only have one and
two neutrons more. In contrast, Ref.~\cite{danysz} reports
$B_{\Lambda\Lambda}$($^{10}_{\Lambda\Lambda}$Be) = 17.7 MeV.

From the results of Table~\ref{tab:results}, we conclude that in order
to explain simultaneously the experimental energy of
$^{\phantom{6}6}_{\Lambda\Lambda}$He, $^{10}_{\Lambda\Lambda}$Be and
$^{13}_{\Lambda\Lambda}$B, RPA effects should be taken into
account. Besides, the RPA resummation leads to a new nuclear density
or A dependence of the $\Lambda\Lambda$ potential in the medium which
notably changes $\Delta B_{\Lambda\Lambda}$ and that provides, taking
into account theoretical and experimental uncertainties, a reasonable
description of the currently accepted masses of these three
$\Lambda\Lambda$ hypernuclei. 

These results we find are in agreement with existing calculations
showing that in-medium corrections are important. In particular we
find that including RPA effects, from the particle-hole excitation
picture provides larger binding energies. This is in agreement with
calculations \cite{saha} showing that, in turn, allows the hyperon
appearance at lower densities in matter.  Additional effects may
happen linked to the hadronic interaction showing its crucial role in
neutron stars. For example, the existence of hyperons has been shown
to generate too soft an equation of state, so that the maximum mass of
neutron stars falls below the mass measured for some compact
objects. Inclusion of three-body effects seem not to help in solving
this issue.  The cooling pattern in a neutron stars seem to be largely
affected by the presence of unpaired hyperons that would induce faster
cooling via direct URCA neutrino processes.  Remarkably, the role of
hyperons shows again important for the transport coefficients. In
particular, in bulk viscosity governing the r-mode instability of
rotating neutron stars \cite{jones}.  Gravitational wave emission is
allowed when the neutron star is unstable with respect to the
r-modes. Future detections of this type of radiation depend, again, on
the microscopic description and strength of the YN interaction.

\begin{table}
\begin{center}
\begin{tabular}{cc|cccc|cccc}
\hline
 &  & \multicolumn{4}{c|}{Without RPA} & \multicolumn{4}{c}{With RPA} \\
 & $B_{\Lambda\Lambda}^{\rm exp}$ & & \multicolumn{3}{c|}{$\Lambda_{\phi\Lambda\Lambda}$ [GeV]} &
& \multicolumn{3}{c}{$\Lambda_{\phi\Lambda\Lambda}$ [GeV]} \\
 &    & no $\phi$ & 1.5 & 2.0 & 2.5 & no $\phi$ &
1.5 & 2.0 & 2.5  \\ 
\hline 
\tstrut
$^{\phantom{6}6}_{\Lambda\Lambda}$He & $6.91\pm 0.13$ \cite{nakazawa10} &6.15 &6.22 &6.53
&6.84 &6.34 & 6.41 &6.82 & 7.33  \\
$^{10}_{\Lambda\Lambda}$Be &$17.7 \pm 0.4$ \cite{danysz}  &
13.1 &13.2 &13.7 &14.2 &14.5 &14.6 &15.6 &16.8  \\
$^{10}_{\Lambda\Lambda}$Be &$11.9 \pm 0.13$ \cite{nakazawa10}  &
     &     &     &     &     &     &     &      \\
$^{13}_{\Lambda\Lambda}$B\phantom{e} & $23.3\pm 0.7$ \cite{nakazawa10}  
&22.5 &22.6 &23.2 &23.8 &24.2 &24.2 &25.4 &27.0  \\
$^{42}_{\Lambda\Lambda}$Ca & $-$ 
&37.2 &37.3 &37.7 &38.1 &38.3 &38.2 &39.1 &40.1  \\
$^{92}_{\Lambda\Lambda}$Zr & $-$ 
&44.1 &44.2 &44.4 &44.7 &44.6 &44.7 &45.2 &46.0  \\
$^{210}_{\Lambda\Lambda}$Pb & $-$ 
&53.1 &53.1 &53.3 &53.4 &53.4 &53.4 &53.7 &54.1  \\
\hline
\end{tabular}
\caption{\label{tab:results} Binding energies $B_{\Lambda\Lambda}$ calculated with our model.}
\end{center}
\end{table}

%
%



\end{document}